# Gapless spin liquid ground state in the $S = \frac{1}{2}$ vanadium oxyfluoride kagome antiferromagnet [NH$_4$]$_2$[C$_7$H$_{14}$N][V$_7$O$_6$F$_{18}$]

L. Clark,[1] J. C. Orain,[2] F. Bert,[2] M. A. de Vries,[1] F. H. Aidoudi,[3] R. E. Morris,[3] P. Lightfoot,[3] J. S. Lord,[4] M. T. F. Telling,[4,5] P. Bonville,[6] J. P. Attfield,[1] P. Mendels[2,7] and A. Harrison[1,8]

[1]CSEC and School of Chemistry, The University of Edinburgh, Edinburgh, EH9 3JZ, United Kingdom
[2]Laboratoire de Physique des Solides, UMR CNRS, Université Paris-Sud, 91504 Orsay, France
[3]School of Chemistry and EaSTChem, University of St Andrews, St Andrews, KY16 9ST, United Kingdom
[4]ISIS Facility, Rutherford Appleton Laboratory, Chilton, Didcot, Oxon, OX110QX, United Kingdom
[5]Department of Materials, University of Oxford, Parks Road, Oxford, OX1 3PH, United Kingdom
[6]Service de Physique de l'État Condensé, CEA-CNRS, CE-Saclay, F-91191 Gif-Sur-Yvette, France
[7]Institut Universitaire de France, 103 bd Saint Michel, F-75005 Paris, France
[8]Institut Laue Langevin, 6 rue Jules Horowitz, 38042 Grenoble Cedex 9, France

The vanadium oxyfluoride [NH$_4$]$_2$[C$_7$H$_{14}$N][V$_7$O$_6$F$_{18}$] (DQVOF) is a geometrically frustrated magnetic bilayer material. The structure consists of $S = \frac{1}{2}$ kagome planes of V$^{4+}$ $d^1$ ions with $S = 1$ V$^{3+}$ $d^2$ ions located between the kagome layers. Muon spin relaxation measurements demonstrate the absence of spin freezing down to 40 mK despite an energy scale of 60 K for antiferromagnetic exchange interactions. From magnetization and heat capacity measurements we conclude that the $S = 1$ spins of the interplane V$^{3+}$ ions are weakly coupled to the kagome layers, such that DQVOF can be viewed as an experimental model for $S = \frac{1}{2}$ kagome physics, and that it displays a gapless spin liquid ground state.

The search for quantum spin liquids (QSLs) has been a major theme in condensed matter research since the proposal by Anderson of the 'resonating valence bond' (RVB) state [1] that can rival the conventional Néel ground state of long range antiferromagnetic order. The RVB state is formed from a superposition of strongly entangled spin singlet pairings whose interactions may be short- (nearest neighbour) or long-range, resonating over the entire spin lattice [2]. The $S = \frac{1}{2}$ kagome antiferromagnet (KAFM), which is composed of corner-sharing equilateral triangles of antiferromagnetically interacting $S = \frac{1}{2}$ ions, is the prime candidate to host this exotic state of matter in two dimensions. This stems from the combination of geometric frustration and strong quantum effects that allow spin fluctuations to persist down to $T = 0$. The exact ground state of the $S = \frac{1}{2}$ KAFM remains controversial as theory predicts several possibilities for the low energy excitation spectra of QSLs. They may be gapped, such as $Z_2$ or short-range RVB liquids, or gapless, with the existence of zero energy spin excitations, which are known as algebraic or long-range RVB liquids [3]. Predictions from density matrix renormalization group algorithms are that the ground state is a fully gapped, topologically ordered $Z_2$ spin liquid, with an estimated spin gap of 0.13(1) $J$ (where $J$ is the antiferromagnetic exchange energy) to deconfined, fractional spinon (spin-½, charge-0) excitations [4,5]. However, other theories expect a gapless spin liquid ground state [6-8], and the best experimental candidate to date, the $S = \frac{1}{2}$ KAFM herbertsmithite, (ZnCu$_3$(OH)$_6$Cl$_2$), appears to be a gapless QSL [9,10].

Experimental realizations of the $S = \frac{1}{2}$ KAFM remain rare and are all based on kagome networks of Cu$^{2+}$ $d^9$ $S = \frac{1}{2}$ cations [11-14]. However, Aidoudi et al. recently reported the synthesis of [NH$_4$]$_2$[C$_7$H$_{14}$N][V$_7$O$_6$F$_{18}$] (diammonium quinuclidinium vanadium(III,IV) oxyfluoride; DQVOF [15] that contains a $S = \frac{1}{2}$ KAFM network of V$^{4+}$ $d^1$ cations and thus provides a new opportunity for the study and understanding of kagome physics, as the role played by interactions such as Jahn-Teller distortions that can govern the low temperature magnetic properties of Cu$^{2+}$ kagome systems [16], may be suppressed here. The trigonal $R-3m$ structure of DQVOF contains two distinct vanadium sites, with V$^{4+}$ $S = \frac{1}{2}$ ions in the kagome layers and V$^{3+}$ $S = 1$ ions between layers. The system was reported to show an absence of long range magnetic order down to 2 K from initial magnetic susceptibility measurements despite evidence for significant antiferromagnetic exchange interactions. In addition, Aidoudi et al. put forward an orbital coupling argument to suggest that the poor V$^{4+}$-V$^{3+}$ superexchange pathway within the structure should result in magnetically isolated $S = \frac{1}{2}$ kagome planes at low temperatures. In this Letter we present further magnetization and susceptibility measurements plus low temperature heat capacity and muon spin relaxation ($\mu$SR) data which demonstrate that DQVOF is a good candidate QSL material with a gapless ground state.

All measurements were performed on a polycrystalline single-phase sample of DQVOF, prepared by the method given in ref. [15]. Magnetic susceptibility was measured in a Quantum Design MPMS SQUID magnetometer from 1.8 K to 300 K. Magnetization data were collected at a temperature of 1.7 K in applied fields up to 14 T in a Cryogenic vibrating sample magnetometer (VSM). The magnetization-field variation is shown in Fig. 1. The net magnetization may be considered as a sum of two components; a Brillouin-like component, which we tentatively attribute to the $V^{3+}$ spins between the kagome layers and a linear contribution, expected from the strongly interacting $V^{4+}$ kagome layers [17,18]. Subtraction of the linear contribution leaves the saturated magnetization of the interlayer $V^{3+}$ spins, which was normalised by $M_{sat} = N_A g \mu_B + 6 N_A g \mu_B /2 = 8 N_A \mu_B$ for $g = 2$, to account for the ideal total of 6 $V^{4+}$ and 1 $V^{3+}$ spins per formula unit of DQVOF. The experimentally determined value of $M/M_{sat} = 0.148(5)$ corresponds to a saturated moment of 1.18 $\mu_B$ for the interlayer $S = 1$ $V^{3+}$ ions. Such moment reductions due to spin-orbit coupling are typical for $V^{3+}$ [19,20], for example $CdV_2O_4$ [21] and $LaVO_3$ [22] have low temperature ordered moments of 1.19 $\mu_B$ and 1.15 $\mu_B$, respectively. In agreement with ref. [15], the susceptibility of DQVOF shows Curie-Weiss behaviour with a Weiss temperature $\theta = -58(4)$ K indicating the dominance of antiferromagnetic exchange interactions (Fig. 1, inset). The small reduction in effective moment, $\mu_{eff} = 4.97(8)$ $\mu_B$ per formula unit, with respect to the expected value of 5.1 $\mu_B$ is again in keeping with $V^{3+}$ spin-orbit coupling.

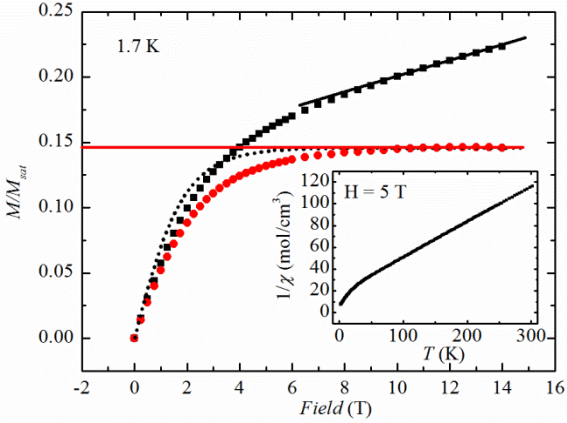

FIG.1. (Color online) Squares: normalised magnetization of DQVOF at 1.7 K with a linear fit to the data above H = 10 T. This response is most likely representative of the non-saturated intrinsic susceptibility of the $V^{4+}$ kagome layers. Circles: the contribution of the $V^{3+}$ spins obtained by subtracting the linear contribution from the total magnetization curve. The dashed curve is the Brillouin function for non-interacting $S = 1$ spins. Inset: inverse susceptibility measured in a 5 T field.

Low temperature heat capacity was measured on a 1.6 mg pressed powder pellet in a Quantum Design PPMS with a $^3$He insert, in zero field (ZF) and applied fields up to 9 T. The data, presented in Fig. 2(a), show a broad Schottky feature that shifts to higher energies in stronger applied fields. This is most likely due to the weakly interacting $S = 1$ spins of the $V^{3+}$ ions located in between the kagome planes. In stronger applied fields (H $\geq$ 5 T) a second feature appears within the low temperature region of the data. We attribute this to the splitting of the $I = \frac{1}{2}$ nuclear spins of the $^1$H and $^{19}$F nuclei that are abundant within the system. The lattice heat capacity and the contribution from the $S = \frac{1}{2}$ spins of the $V^{4+}$ ions within the kagome layers are most likely field independent at these field strengths, given the relatively strong antiferromagnetic interaction [23,24]. In order to isolate the field dependent contributions, the zero field heat capacity was subtracted from the applied field data, for example, the difference between 9 T and zero field heat capacities is shown as $\Delta C_v T^{-1}$ in the left-hand inset of Fig. 2(a). These difference curves were all successfully fitted by a model consisting of two Schottky anomalies. The nuclear Schottky term corresponding to the 40 hydrogen and fluorine nuclei per formula unit has no free parameters. The electronic Schottky contribution was fitted well by the expression for an $S = 1$ system;

$$C_v T^{-1} = \frac{f_{S=1} k_B N_A}{T} \left(\frac{E_{S=1}}{T}\right)^2 \frac{exp(E_{S=1}/T)+exp(-E_{S=1}/T)+4}{\left(1+exp(E_{S=1}/T)+exp(-E_{S=1}/T)\right)^2} \quad (1)$$

The fitted value of $f_{S=1} = 0.5(2)$ for the fraction of $S = 1$ spins per formula unit shows a similar reduction to the magnetization, as a result of spin-orbit coupling. The obtained Zeeman splitting energy ($E_{S=1} = g\mu_B H$) shown in the right-hand inset of Fig. 2(a), gives $g = 1.8(2)$, which is typical for $V^{3+}$ ions. We note the existence of a small (~0.7 K) zero field splitting of the $S = 1$ levels, which is also likely to result from spin-orbit coupling in the $t_{2g}^2$ $V^{3+}$ ground state. The magnetization and heat capacity results show that the interlayer $V^{3+}$ ions behave as spin-orbit coupled but non-interacting paramagnetic spins, and that to a good approximation DQVOF contains magnetically isolated kagome layers of strongly correlated $S = \frac{1}{2}$ spins of $V^{4+}$ ions.

The field dependent Schottky features were subtracted from the heat capacity data to give, in Fig. 2(b), the contribution of the strongly correlated spins within the kagome layers together with the lattice contribution, which is expected to be vanishingly small at low temperatures. The observation that all subtracted curves approximately coincide reflects the quality of the fit of the field dependant heat capacity to our model. It also implies that the low temperature ground state of DQVOF is robust to the application of fields up to 9 T, and so forms a quantum critical phase rather than sitting at a quantum critical point. The data show no sharp peak structure within the measured temperature range indicating the absence of any magnetic phase transitions. This is in agreement with the magnetization data and the $\mu$SR measurements discussed below. In the absence of a non-magnetic analogue, the phonon contribution cannot be subtracted accurately but an upper bound of the lattice contribution is estimated by a $T^3$ phonon contribution that matches the total heat capacity at 30 K. This lattice contribution is found to be negligible below 5 K. A recent estimate for the spin gap in a topologically ordered spin liquid [4,5] is 0.13$J$ ≈ 7.5 K for DQVOF, taking the nearest neighbor exchange energy as $J = -\theta$ from the mean-field description of the $S = \frac{1}{2}$ KAFM. However, the continuous density of states observed down to 300 mK implies that there is no complete spin excitation gap in the spectrum of DQVOF within the measured energy range. Various models have been proposed for the low temperature heat capacity of gapless spin liquids. Algebraic or Dirac spin liquids are expected to show a $C_v \propto T^2$ dependence at low temperature [6,7]. This does not appear to be applicable to DQVOF. A better description seems to be given by a more conventional spinon Fermi surface with a $T$-linear behaviour in the low temperature specific heat. Modelling the data with $C_v = \gamma T$ between 300 mK and 5 K gives a reasonable fit to the data with $\gamma \sim$ 200 mJ K$^{-2}$ mol$^{-1}$ per V$^{4+}$ spin. $\gamma$ is related to the spinon density of states at the spinon Fermi surface, and the value found here to be comparable to those from other experimental gapless spin liquids [25-27]. However, the fit in Fig. 2(b)) is improved by using a $C_v = \gamma T^\alpha$ term with exponent $\alpha = 1.2$. This is remarkably similar to the behavior of herbertsmithite where the low temperature heat capacity data was modelled with a $T^{1.3}$-dependence [24].

A key feature of a vanishing spin gap in a QSL is the absence of spin freezing and persistence of internal field fluctuations down to $T = 0$. Experimentally, this can be observed by $\mu$SR due to the extreme sensitivity of the muon to small magnetic moments. Measurements were performed on the MuSR spectrometer at the ISIS pulsed muon facility, Rutherford Appleton Laboratory, U.K. The background fraction of muons landing in the sample backing plate or elsewhere was determined by comparison of the data with those taken on the GPS instrument at the Laboratory for Muon Spin Spectroscopy, Paul Scherrer Institute, Switzerland at the same temperature and magnetic field. The PSI equipment used an active veto detector to give very low background [28]. Figure 3 shows the time dependence of the background corrected muon spin polarization in zero field (ZF) and an applied longitudinal fields (LF) of 200 G at 40 mK and 40 K.

The ZF depolarization reflects the sum of the local responses of muons implanted at different stopping sites within the sample. Implanted muons are likely to be at sites near oxide or fluoride anions, due to their favourable electrostatic interactions with the positive muons. A fit to the high temperature ZF depolarization data [29] reveals that the kink at 2 $\mu$s can be ascribed to a strong dipolar coupling between the muon spin and the fluorine $I = \frac{1}{2}$ nuclear spin [30,31]. This fluorine response accounts for 10% of the muons stopping within the sample. The presence of this characteristic feature in the ZF data at all temperatures indicates that the muons implanted at the fluorine site are only very weakly coupled to the electronic spins of the system [29]. In addition, a small fraction of muons most likely stop far from the kagome planes and give rise to a small non-relaxing polarization, ~ 0.1, clearly visible at long time. The remainder of the implanted muons fall at sites situated near oxide anions.

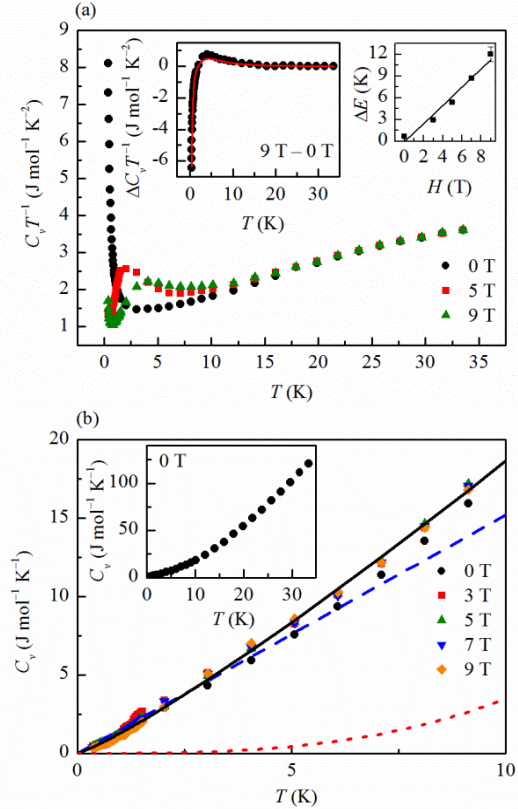

FIG. 2. (Color online) (a) $C_vT^{-1}$ of DQVOF against temperature measured in ZF and applied fields of 5 T and 9 T. Left-hand inset the fit to the Schottky anomaly (solid line) to the difference of interpolated heat capacity in 9 T and 0 T. The right-hand inset shows the Zeeman splitting of the $S = 1$ levels. (b) The low temperature heat capacity after subtraction of the Schottky anomaly with $C_v = \gamma T$ (dashed line) and $C_v = \gamma T^{1.2}$ (solid line) models. The dotted red curve gives an estimate of the lattice contribution to the heat capacity. Inset shows the ZF data over the entire measured temperature range.

Both the 40 K and 40 mK data in Fig. 3 relax to the same value of polarization pointing to the absence of a '⅓-tail' component characteristic of a magnetically-frozen state at long times in the 40 mK data. This precludes any argument for static magnetism in the sample upon cooling [32]. Hence the moderate increase of the relaxation upon cooling from high temperature reflects only a slowing down of the electronic spin dynamics. The local nuclear fields at each stopping site are static on the timescale of the muon experiment and can be decoupled from the muon upon application of a LF of 200 G. As a result, the muon spin only probes the dynamical, temperature dependent field fluctuations arising from the electronic spins within the system, which are of principal interest here. The time evolution of the spin depolarization in an applied LF of 200 G was modelled with a stretched exponential relaxation function over the entire measured temperature range,

$$P(t) = P_{non-relaxing} + P_O \exp(-\lambda t)^{\beta} \quad (2)$$

Figures 4(a) and (b) show the temperature dependence of the spin depolarization rate, $\lambda$, and the stretching component, $\beta$, respectively.

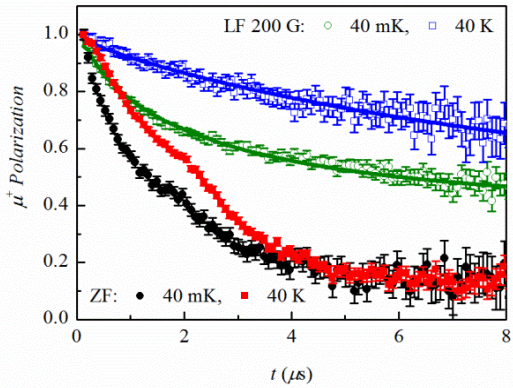

FIG. 3. (Color online) Time dependent muon spin polarization of DQVOF at 40 mK and 40 K in zero field (ZF) and an applied longitudinal field (LF) of 200 G. Solid lines are fits to the data (Equation 2).

Above 10 K, the system is in a fast fluctuating regime from which there is a slowing of spin dynamics down to 1 K. Below 1 K the data show a plateau in the depolarization rate; a clear indicator of persistent dynamics at low temperatures. The limiting value of $\beta$ ~0.5 implies a distribution of dynamical correlations at low temperatures and the muons implanted at the oxygen site are most likely under the influence of more than one fluctuation frequency. The origin of the muon spin relaxation in DQVOF requires some consideration given that there are two potential sources of electronic field fluctuations, namely, the $V^{3+}$ $S = 1$ and the $V^{4+}$ $S = ½$ spins. We estimate that the relaxation rate, $\lambda$, due to the $S = 1$ spins of $V^{3+}$ is ~0.01 MHz, from the effective interaction between $S = 1$ spins of ~1 K as observed from the magnetization against field data shown in Fig. 1 [29]. Given that the experimentally determined value is larger than this, it implies that the slowing down of electronic spin dynamics and the relaxation shown in Figure 4(a) is intrinsic to the kagome layers. The persistence of electronic field fluctuations within the kagome layers at the lowest measured temperatures, therefore, supports the existence of a gapless spin liquid state in DQVOF as deduced from the low temperature heat capacity data.

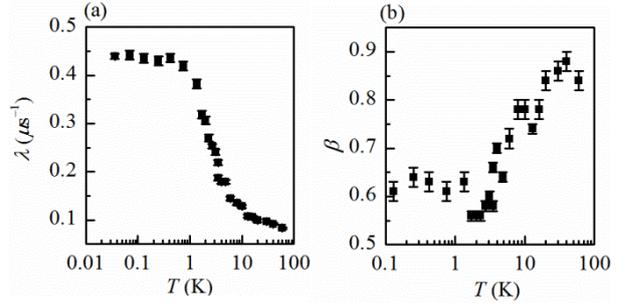

FIG. 4 The temperature dependence of (a) the muon spin depolarization rate, $\lambda$, and (b) the stretching component, $\beta$, of DQVOF.

In summary, this study of low temperature magnetism in DQVOF demonstrates that the interplane $V^{3+}$ $S = 1$ spins show paramagnetic single-ion ion behaviour with significant spin-orbit coupling effects, while the $V^{4+}$ layers are a good experimental realization of an $S = ½$ kagome antiferromagnet. $\mu$SR shows that there is no spin freezing within DQVOF down to 40 mK. Such persistent spin dynamics at low temperatures are a key requirement for a QSL state. The low temperature heat capacity of the $V^{4+}$ kagome planes displays a continuous density of magnetic states down to 300 mK demonstrating that there is no complete spin gap in the low energy excitation spectrum, and shows little variation in fields up to 9 T, consistent with a gapless quantum critical phase spin liquid ground state. This heat capacity varies as $T^{1.2}$ which is remarkably similar to the $T^{1.3}$-dependence of herbertsmithite [24]. DQVOF joins just two $Cu^{2+}$ materials, herbertsmithite and vesignieite, as experimental realizations of the $S = ½$ kagome lattice. None of these materials show the theoretically favoured gapped spin liquid behaviour of the pure Heisenberg KAFM model [4,5]. Vesignieite orders at a fairly high temperature [33], while both herbertsmithite and DQVOF are found to be gapless spin liquids. It is also notable that a significant, although not critical, slowing of the spin dynamics is detected in DQVOF at low temperatures, but not in herbertsmithite [18]. This suggests that the quantum critical phase in DQVOF lies closer to some uncovered phase transition than in the case of herbertsmithite. Such a phase transition will be driven by perturbations that will differ in magnitude between $d^1$ and $d^9$ systems, for example the spin-orbit driven Dzyaloshinskii-Moriya interaction. It will be thus be particularly interesting to discover possible instabilities of the liquid ground state of DQVOF by applying stronger magnetic fields [34] or pressure [35] and hence explore the perturbations that can pin the ground state of these kagome models.

This work was supported in part by the Grant No. ANR-09-JCJC-0093-01, and by EPSRC, STFC and the Royal Society. We thank A. Amato for assistance at PSI.